\documentclass[reprint,amsmath,amssymb,aps,pra]{revtex4-2}
\usepackage{graphicx}
\usepackage{dcolumn}
\usepackage{bm}
\usepackage[hidelinks]{hyperref}
\hypersetup{colorlinks=true, urlcolor=blue, linkcolor=blue, citecolor=blue}

\begin{document}

\title{Criticality and Chaos in Auditory and Vestibular Sensing}
\author{Justin Faber$^1$}
\email{faber@physics.ucla.edu}
\author{Dolores Bozovic$^{1, 2}$}
\email{bozovic@physics.ucla.edu}
\affiliation{ $^1$Department of Physics \& Astronomy and $^2$California NanoSystems Institute, University of California, Los Angeles, California 90095, USA\\}
\date{\today}

\begin{abstract}
\noindent \textbf{The auditory and vestibular systems exhibit remarkable sensitivity of detection, responding to deflections on the order of Angstroms, even in the presence of biological noise. Further, these complex systems exhibit high temporal acuity and frequency selectivity, allowing us to make sense of the acoustic world around us. As this acoustic environment of interest spans several orders of magnitude in both amplitude and frequency, these systems rely heavily on nonlinearities and power-law scaling. The behavior of these sensory systems has been extensively studied in the context of dynamical systems theory, with many empirical phenomena described by critical dynamics. Other phenomena have been explained by systems in the chaotic regime, where weak perturbations drastically impact the future state of the system. We first review the conceptual framework behind these two types of detectors, as well as the detection features that they can capture. We then explore the intersection of the two types of systems and propose ideal parameter regimes for auditory and vestibular systems.}
\end{abstract}

\maketitle

\section{Introduction and Overview}

Auditory and vestibular systems perform tasks that are crucial for the survival of an animal, enabling it to navigate in space, detect signals from predators or prey, identify and attract potential mates, and communicate with members of the same species. To achieve these tasks, the sensory system must detect extremely weak signals, extract them from noisy environments, distinguish tones of closely spaced frequencies, and precisely parse temporal information. Specifically, near our threshold of hearing, we are able to detect displacements of the eardrum smaller than the width of the hydrogen atom  \cite{dalhoffDistortionProductOtoacoustic2007a}. This detection occurs in the presence of internal thermal fluctuations of equal or higher magnitude. Further, temporal resolution of humans typically reaches 10 microseconds \cite{leshowitzMeasurementTwoClick2005a, brugheraHumanInterauralTime2013}, enabling sound localization through interaural time differences \cite{grotheMechanismsSoundLocalization2010}. Finally, we are able to resolve tones that differ in frequency by only a fraction of a percent \cite{hudspethIntegratingActiveProcess2014}. In parallel, the auditory system achieves immense dynamic range in both amplitude and frequency of acoustic detection. We are able to detect sounds that span over 12 orders of magnitude in intensity and 3 orders of magnitude in frequency. These broad ranges are reflected in our logarithmic decibel scale for sound intensity and logarithmic spacing of musical intervals.

The remarkable features of the auditory system rely heavily on two factors: active amplification and nonlinear response. First, the system has been shown to violate the fluctuation dissipation theorem, indicating that it cannot be governed by equilibrium statistical mechanics \cite{Martin2001}. Vast amounts of empirical evidence obtained \textit{in vivo} demonstrate that the inner ear expends energy to amplify signals \cite{Hudspeth2008}. Second, compressive nonlinearities in the response to external stimuli enable the extensive dynamic range, while maintaining sensitivity to weak signals \cite{eguiluzEssentialNonlinearitiesHearing2000, Martin2001}. These nonlinearities have been detected at all scales measured, from individual sensory cells to \textit{in vivo} phenomena known as phantom tones \cite{TartiniRef, jaramilloAuditoryIllusionsSingle1993, barralPhantomTonesSuppressive2012a}. How the auditory system utilizes energy expenditure and nonlinearities to achieve its remarkable detection characteristics, however, remains unknown after more than 7 decades of research \cite{Gold1, Gold2, Reichenbach2014}.

The theoretical framework for auditory detection that was developed over the past 20 years is based on the notion of a dynamical system poised near criticality \cite{dukeCriticalOscillatorsActive2008}, on the verge of autonomous oscillation. The models apply the normal form equation for the supercritical Hopf bifurcation to describe the auditory system, and elegantly capture the mechanical sensitivity, frequency selectivity, and amplitude-compressive response, thus reproducing a broad range of experimental results \cite{martinCompressiveNonlinearityHair2001, roblesMechanicsMammalianCochlea2001}. In the vicinity of the bifurcation, the system's sensitivity to external signals increases, while the frequency selectivity sharpens, thus pointing to criticality as the optimal regime for signal detection \cite{choeModelAmplificationHairbundle1998}. 
 
However, while proximity to a bifurcation yields many advantages, the description contains an innate constraint: at criticality, the system becomes infinitely slow, with transient times diverging as a result of \textit{critical slowing down}. This sluggish behavior poses an undesirable tradeoff between sensitivity of a detector and its speed, and is inconsistent with the high temporal acuity exhibited by our auditory system. Furthermore, at the critical point, the system is very sensitive to the effects of stochastic fluctuations, which limits some of the advantages observed in the deterministic models. As noise is a ubiquitous component of any biological system, and its effects specifically near criticality are not negligible, this constraint limits the advantages of tuning a system near a bifurcation point. 

An alternate theoretical framework, developed to reconcile the requirement for high sensitivity of detection with the need for a rapid temporal response, is based on the notion of chaotic dynamics \cite{faberChaoticDynamicsEnhance2019a}. When a system is poised in a chaotic regime, even infinitesimally weak external perturbations trigger large changes in the subsequent trajectory, thus yielding high sensitivity. As a result of this exponential divergence, the system also responds and resets rapidly. Hence, a chaotic system avoids the inherent tradeoff observed with criticality. In a prior study, we demonstrated experimentally the presence of a chaotic attractor in the innate and driven oscillations exhibited by sensory cells \textit{in vitro} \cite{faberChaoticDynamicsInner2018}. Using mathematical methods from dynamical systems literature, we confirmed that the oscillator contains a deterministic component and is not completely dominated by biological noise and other stochastic processes. Further, we showed chaos  to be beneficial in this system, as the instabilities from which chaos arises enhance the sensitivity and temporal resolution of the response \cite{faberChaoticDynamicsEnhance2019a}. 

We note that chaos is typically considered a harmful element in applied mathematics literature, as it limits control and predictability. However, it has been proposed to play a potentially helpful role in certain biological systems, as it enhances their dynamical complexity \cite{kaplanSubthresholdDynamicsPeriodically1996, obyrneHowCriticalBrain2022}. As chaotic regimes can arise in the presence of three or more degrees of freedom, we predict that many more examples of living systems utilizing chaotic dynamics will be uncovered in the future.

In the present work, we aim to compare the relative advantages of chaos versus criticality in the theoretical description of auditory and vestibular detection. In particular, we explore the interface of these two theoretical models: a system poised near the supercritical Hopf bifurcation and one poised in the chaotic regime, as well as the continuum describing transitions between the regimes. To assess separately each of the key features exhibited by these sensory systems, in each section, we characterize the sensitivity, frequency selectivity, temporal acuity, and power-law amplitude response of a Hopf oscillator, near and far from criticality, in the presence and absence of chaos. As stochastic fluctuations play a non-negligible role in dynamical systems, we find the parameter regimes that optimize these detection characteristics in the presence of noise. We then combine these independent metrics to propose a simple conceptual framework for auditory and vestibular detection.

\begin{figure*}[t]
\centering
\includegraphics[width=\textwidth]{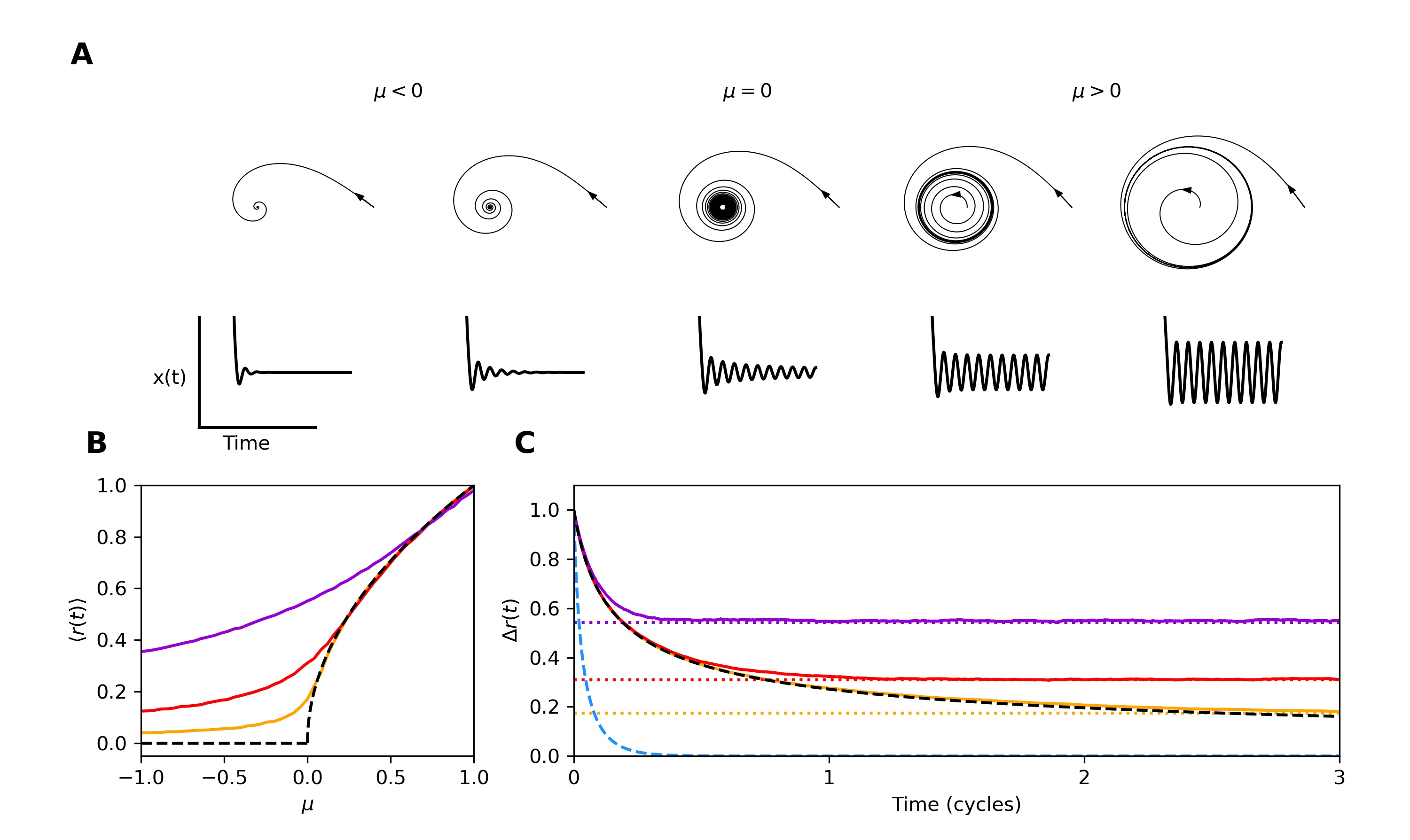} 
\caption{(A) State-space dynamics (top row) and time traces (bottom row) of the Hopf oscillator for several values of the control parameter, increasing from left to right. (B) steady-state mean amplitude of the system near the bifurcation for several levels of additive noise. (C) Return to steady-state following a mechanical perturbation to the critical system ($\mu=0$) averaged over many trials for several noise levels. Black, orange, red, and purple correspond to $D=0, 0.001, 0.01$, and $0.1$, respectively. The blue, dashed curve shows a comparison to a system with $\mu=1$ and $D=0$. Dotted lines indicated the steady-state mean amplitude for each noise level. For the deterministic cases, this value is zero.}
\label{fig:Hopf}
\end{figure*}

\begin{figure*}[t]
\centering
\includegraphics[width=\textwidth]{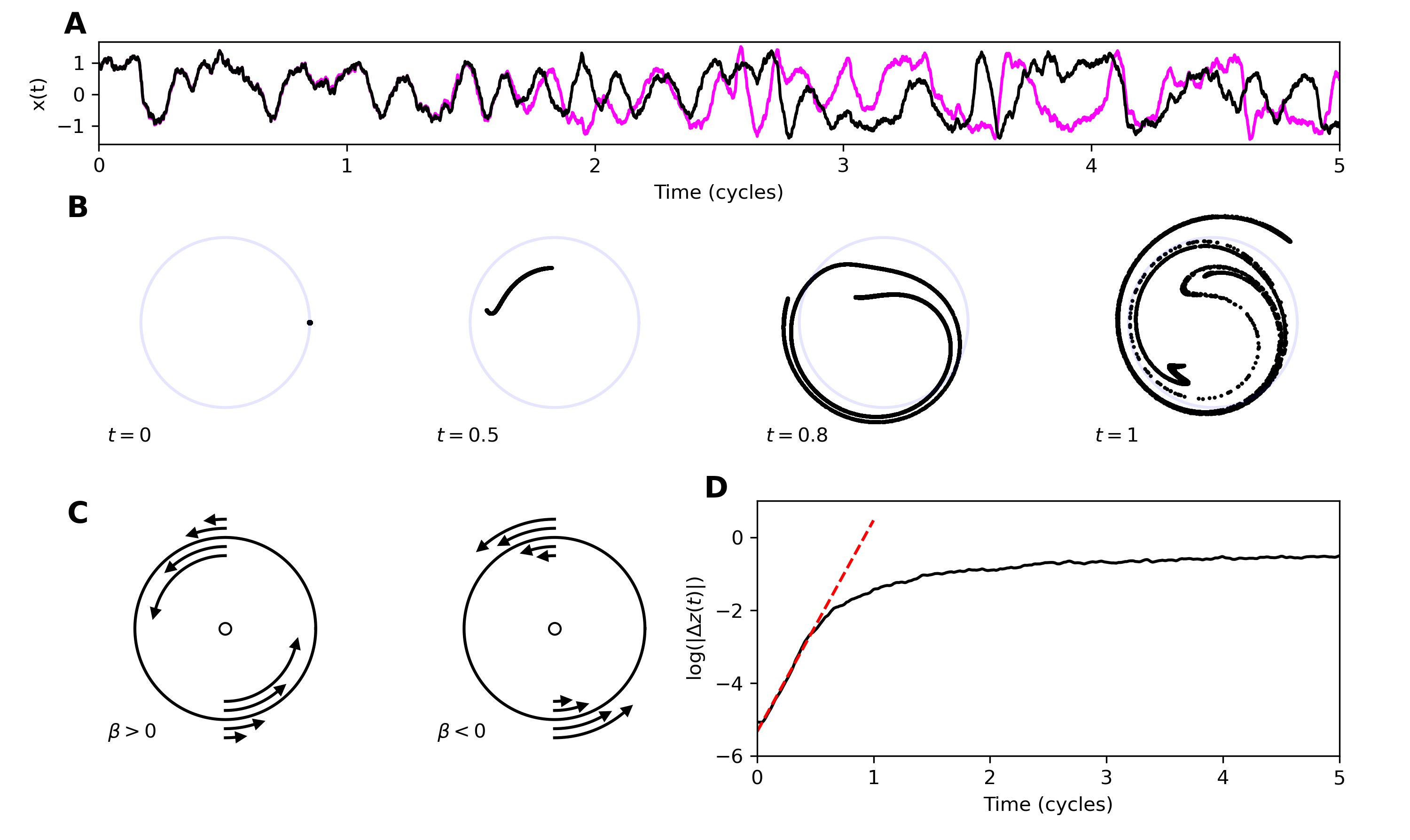}
\caption{(A) Two diverging time series traces of the nonisochronous Hopf oscillator given identical realizations of noise but slightly different initial conditions. (B) Snapshots at four points in time of $10^4$ systems prepared all with slightly different initial conditions. The circles illustrate the location of the limit cycle in the absence of noise. (C) Illustrations of the phase space vector field of nonisochronous oscillators. Angular velocity (frequency) is dependent on the radius (amplitude) of oscillation. (D) Natural log of the average difference between pairs of trajectories, averaged over many realizations of noise. The initial linear growth corresponds to exponential divergence of trajectories. The slope of the best-fit line to the initial divergence (red-dashed) corresponds to the Lyapunov exponent.}
\label{fig:Noniso}
\end{figure*}

\section{Hopf bifurcation}

The inner ear of vertebrates contains a number of end organs that specialize in either auditory or vestibular detection, the latter including both translational and rotational movement. While the pathways by which external signals reach the inner ear vary, they ultimately result in mechanical vibrations of internal structures; hence, the two sensory systems exhibit many features in common. Conversion of mechanical energy of a sound, vibration, or acceleration into electrical energy in the form of ionic currents is performed by specialized, sensory hair cells of both the auditory and vestibular systems. The hair cell gets its name from the rod-like, inter-connected stereocilia that protrude from the apical surface, which are collectively named the hair bundle. An incoming stimulus pivots the stereocilia and modulates the open probability of the transduction channels embedded in the tips of the stereocilia \cite{lemasurierHairCellMechanotransductionCochlear2005, vollrathMicromachineryMechanotransductionHair2007a}. In several species, these hair bundles have been shown to exhibit active limit-cycle oscillations in the absence of applied stimulus \cite{Martin2003, benserRapidActiveHair1996, crawfordMechanicalPropertiesCiliary1985,Martin2001}. Though the role of these spontaneous hair-bundle oscillations \textit{in vivo} has yet to be established, they serve as an experimental probe for studying this active system, as they lead to sub-nanometer thresholds \textit{in vitro} \cite{narinsVertebrateEarExquisite1984, ActiveHairbundleMovements1999}.

The Hopf oscillator has been extensively used for modeling and understanding the phenomenon of active, spontaneous hair-bundle oscillations, as well as more global features of the auditory and vestibular systems \cite{choeModelAmplificationHairbundle1998}. When the system is poised on the verge of instability (near the Hopf bifurcation), it becomes extremely sensitive to small perturbations. In the absence of noise, the amplitude gain of the response diverges as the system approaches criticality. The model was shown to capture active amplification and power-law amplitude response observed both \textit{in vivo} and \textit{in vitro} on a number of different species \cite{hudspethIntegratingActiveProcess2014}.

A Hopf oscillator is described by time-dependent complex variable, $z(t)$, which is governed by a normal form equation, which in its simplest version, takes the form:

\begin{eqnarray}
\frac{dz}{dt} = (\mu + i\omega_0)z - |z|^2z + F(t) + \eta(t)
\label{eq:hopf1}
\end{eqnarray}

\noindent where $\mu$ and $\omega_0$ are the control parameter and characteristic frequency of the detector, respectively. $F(t)$ represents the external forcing on the system, while $\eta(t)$ is a stochastic variable, representing thermal noise. This variable is complex, with independent real and imaginary parts, both of which have statistics of Gaussian white noise: $\langle\eta(t)\rangle = 0$, $\langle\eta(t)\eta(t')\rangle = 0$, and $\langle\eta(t)\bar{\eta}(t')\rangle = 4D\delta(t-t')$, where $\bar{\eta}$ is the complex conjugate of $\eta$ and $D$ defines the noise strength.

In the absence of forcing and noise ($F(t) = D = 0$), the system can be more easily understood in polar coordinates, by letting $z(t) = r(t)e^{i\theta(t)}$, thereby separating the complex variable into two real variables. This results in the pair of equations,

\begin{eqnarray}
\frac{dr}{dt} = \mu r - r^3  \text{  \quad and  \quad  } \frac{d\theta}{dt} = \omega_0,
\label{eq:hopf1b}
\end{eqnarray}

\noindent which describe the amplitude and phase dynamics of the system, respectively. Notice that the instantaneous frequency $\frac{d\theta}{dt}$ is constant, having no dependence on the oscillator's amplitude, $r(t)$. This defines an isochronous oscillator. 

The amplitude dynamics are determined by the control parameter, $\mu$. For $\mu < 0$, the system displays a stable fixed point with increasing stability for more negative values of the control parameter. As the control parameter approaches the critical point at $\mu = 0$, the system loses stability and spontaneous limit-cycle oscillations emerge. For $\mu > 0$, as this parameter increases in the positive regime, the spontaneous oscillations grow larger and more stable. Precisely at the critical point, the system becomes infinitely sensitive in the absence of noise, as an infinitesimal perturbation can cause large amplitude displacements. For both positive and negative values of the control parameter, the amplitude returns to its steady-state exponentially, with characteristic time scale proportional to $1/\mu$ (see the Appendix). In the vicinity of the bifurcation, the linear term becomes vanishingly small, and the system returns to steady-state very slowly, with perturbations diminishing according to a power-law (Fig. \ref{fig:Hopf}A).  As the control parameter approaches the Hopf bifurcation, this time scale diverges without bound at $\mu=0$, a dynamical systems phenomenon known as \textit{critical slowing down}. 

In the presence of noise, a system near the critical point becomes very susceptible to stochastic fluctuations. In Fig. \ref{fig:Hopf}B, we show how the amplitude of limit cycles oscillations of the system depend on the control parameter, at different levels of additive noise. As can be seen, the infinite sensitivity that is present in the deterministic limit is removed by the presence of noise, and the bifurcation point is smeared. Noise also obscures or even removes the effects of critical slowing down. We see that the long transient associated with power-law decay at the critical point is shortened, as the steady-state amplitude is no longer zero (Fig. \ref{fig:Hopf}C). The level of noise required to obscure the effect ($D\approx0.1$) significantly smears out the bifurcation diagram, removing the critical point.  Different levels of noise therefore result in different tradeoffs between sensitivity and speed of the response.

\begin{figure*}[t]
\centering
\includegraphics[width=\textwidth]{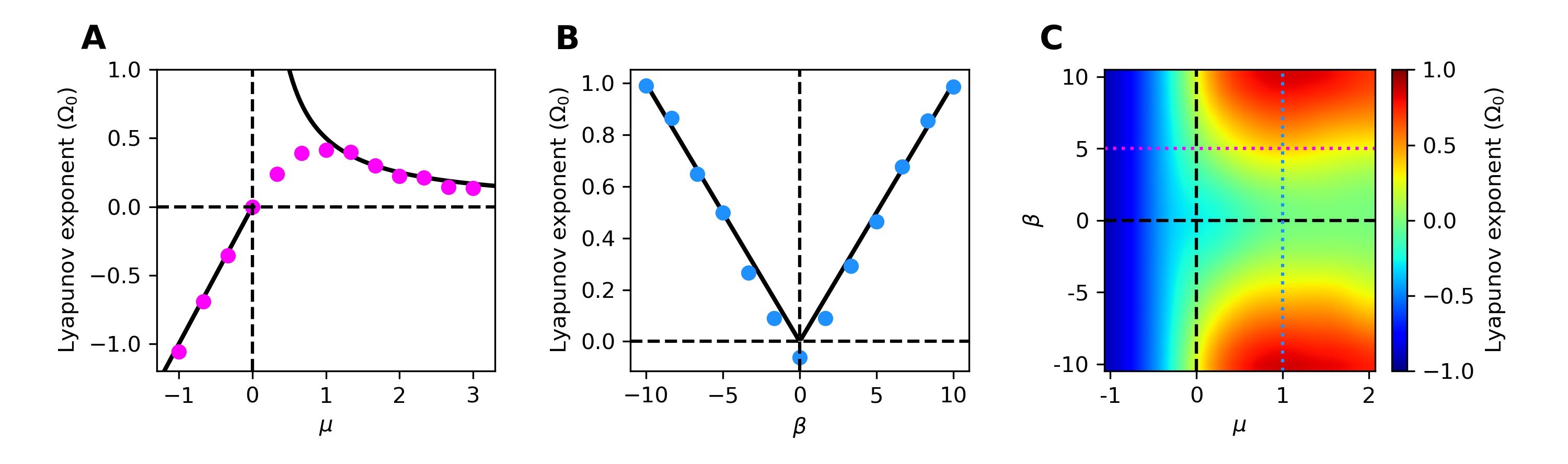}
\caption{(A) Lyapunov exponent calculated numerically with $\beta=5$ for a range of control parameters. (B) Lyapunov exponent calculated numerically with $\mu=1$ for a range of degrees of nonisochronicity. For (A-B), the black curves correspond to the analytic approximations of the Lyapunov exponent. (C) Lyapunov exponent calculated numerically throughout the parameter space. The dotted lines correspond to the cross-sections plotted in (A-B). For all panels, $D=0.1$.}
\label{fig:Lyap}
\end{figure*}

\begin{figure*}[t]
\centering
\includegraphics[width=\textwidth]{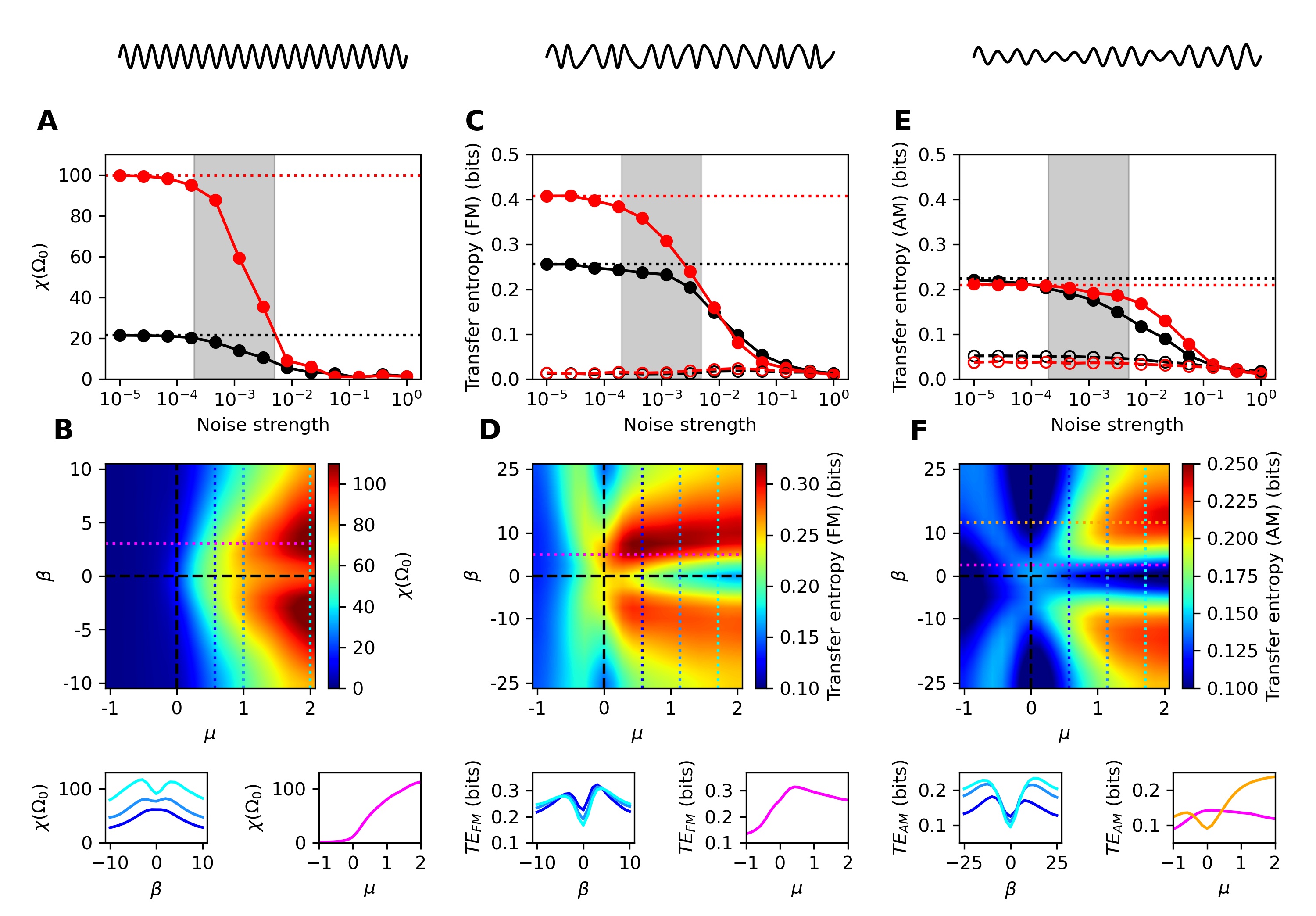}
\caption{(A, C, E) Sensitivity metrics for the critical system ($\mu =\beta = 0$) and chaotic system ($\mu =1, \beta = 5$) for a range of noise strengths. Dotted lines correspond to the value of each metric in the deterministic system. In (C, E), open points correspond to transfer entropy in the reverse direction (response to stimulus) and serve as controls. The grey, shaded regions estimate the biological noise level experienced by hair bundles (1--5\% of the signal amplitude \cite{fredrickson-hemsingDynamicsFreelyOscillating2012}). Stimulus waveforms are illustrated above each column. (B, D, F) Heatmaps of the three measures throughout the parameter space with $D=10^{-3}$. Cross-sections along the colored, dotted lines are shown below.}
\label{fig:Sensitivity}
\end{figure*}

\section{Nonisochronicity and Chaos}

In several studies of auditory and vestibular systems, a more general version of the Hopf oscillator was considered \cite{roongthumskulDynamicsMechanicallyCoupled2021, faberChaoticDynamicsEnhance2019a}. The equation takes the form

\begin{eqnarray}
\frac{dz}{dt} = (\mu + i\omega_0)z - (\alpha + i\beta)|z|^2z + F(t) + \eta(t)
\label{eq:hopf2},
\end{eqnarray}

\noindent where $\alpha$ and $\beta$ are introduced to characterize the nonlinearity of the system, while all the other parameters carry the same meaning as in the isochronous case previously described. In the absence of stimulus and noise ($F(t) = D = 0$), this equation can be written in polar coordinates as

\begin{eqnarray}
\frac{dr}{dt} = \mu r - \alpha r^3  \text{  \quad and  \quad  } \frac{d\theta}{dt} = \omega_0 - \beta r^2
\label{eq:hopf2b}
\end{eqnarray}

\noindent For $\mu > 0$, a stable limit cycle exists at radius $r_0 = \sqrt{\frac{\mu}{\alpha}}$. The frequency at this limit cycle is $\Omega_0 = \omega_0 - \beta r_0^2 = \omega_0 - \beta\mu / \alpha$, where $\omega_0$ is the frequency at the Hopf bifurcation. This more general description reduces to the traditional, isochronous form when $\alpha = 1$ and $\beta = 0$. We restrict our analysis to systems with $\alpha = 1$ and vary $\beta$ to control the level of nonisochronicity. This new parameter causes the frequency of oscillation to depend on the amplitude of oscillation (Fig. \ref{fig:Noniso}C).

By coupling the radial and phase degrees of freedom, the presence of nonisochronicity ($\beta \neq 0$) introduces significant complexity to the dynamics of this system. It distorts the Lorentzian shape of the frequency response curve \cite{gleesonNonLorentzianSpectralLineshapes2006} and can even produce branching of this curve, discontinuities, and bistability for a range of frequencies, as well as hysteretic behavior in response to frequency sweeps \cite{zhangPeriodicallyForcedHopf2011a}. The presence of nonisochronicity also makes the system susceptible to chaotic dynamics. Chaos in dynamical systems is characterized by extreme sensitivity to initial conditions and exponential divergence of neighboring trajectories. In the nonisochronous Hopf oscillator, chaos can arise from sinusoidal or impulsive external forcing \cite{faberChaoticDynamicsEnhance2019a, faberNoiseinducedChaosSignal2019}, as well as systems driven purely by stochastic white noise. To distinguish exponentially diverging, chaotic trajectories from simple diffusion induced by noise, it is useful to introduce identical realizations of noise (common noise) to systems with slightly different initial conditions (Fig. \ref{fig:Noniso}A). If the distance between neighboring trajectories diminishes with time or remains the same, the system is non-chaotic. However, if the common noise stimulus causes neighboring trajectories to diverge exponentially, then the system exhibits noise-induced chaos \cite{Goldobin2005, Neiman2011}. We note that the exponential divergence is observed only over short time scales; as a chaotic system is bounded in phase space, the distance between trajectories plateaus with time (Fig. \ref{fig:Noniso}D).

The rate of divergence of neighboring trajectories is expected to follow $|\Delta z(t)| \propto e^{\lambda t}$, where $\Delta z(t)$ is the Euclidean distance between two nearby phase-space trajectories, and $\lambda$ is the Lyapunov exponent. Chaotic systems are often defined by $\lambda > 0$. For quiescent, non-chaotic systems, the Lyapunov exponent is negative and characterizes the system's stability, or the rate at which the system returns to equilibrium following a perturbation. For a Hopf oscillator with $\mu < 0$ and no forcing or noise, the Lyapunov exponent can be found by expanding around the stable fixed point (see the Appendix). One finds that in this simple case, $\lambda = \mu$, which indicates that the system becomes more stable for more negative values of $\mu$. 

In the absence of noise, as the control parameter crosses from the quiescent to the oscillatory regime, the Lyapunov exponent increases to zero and remains there even for ($\mu > 0$). Since perturbations in the phase neither diverge nor converge, trajectories are neutrally stable along the angular direction of the limit cycle. In the presence of noise-induced chaos, analytic approximations often becomes intractable, but numerical simulations yield values of the Lyapunov exponent that can be positive, negative, or zero. For a noisy nonisochronous Hopf oscillator, an analytic approximation can be made in the regime of sufficiently stable limit cycles. In this regime, the Lyapunov exponent shows a simple dependence on both the control and nonisochronicity parameters \cite{faberChaoticDynamicsEnhance2019a},

\begin{eqnarray}
\lambda \approx \frac{|\beta|D}{\mu}
\label{eq:lyap}.
\end{eqnarray}

\noindent This approximation is particularly useful in weakly chaotic regimes, where numerical simulations are computationally expensive. We show the robustness of this approximation for sufficiently stable limit cycles, as well as the simpler approximation for quiescent systems. The analytic approximation however breaks down as the system approaches criticality from the oscillatory side, because the assumption of a sufficiently stable limit cycle is no longer valid (Fig. \ref{fig:Lyap}A). We provide a map of the Lyapunov exponent calculated numerically throughout the parameter space (Fig. \ref{fig:Lyap}C). These calculations serve to show the degree of chaos as the system crosses the Hopf bifurcation, a regime not explored analytically.

\begin{figure*}[t]
\centering
\includegraphics[width=\textwidth]{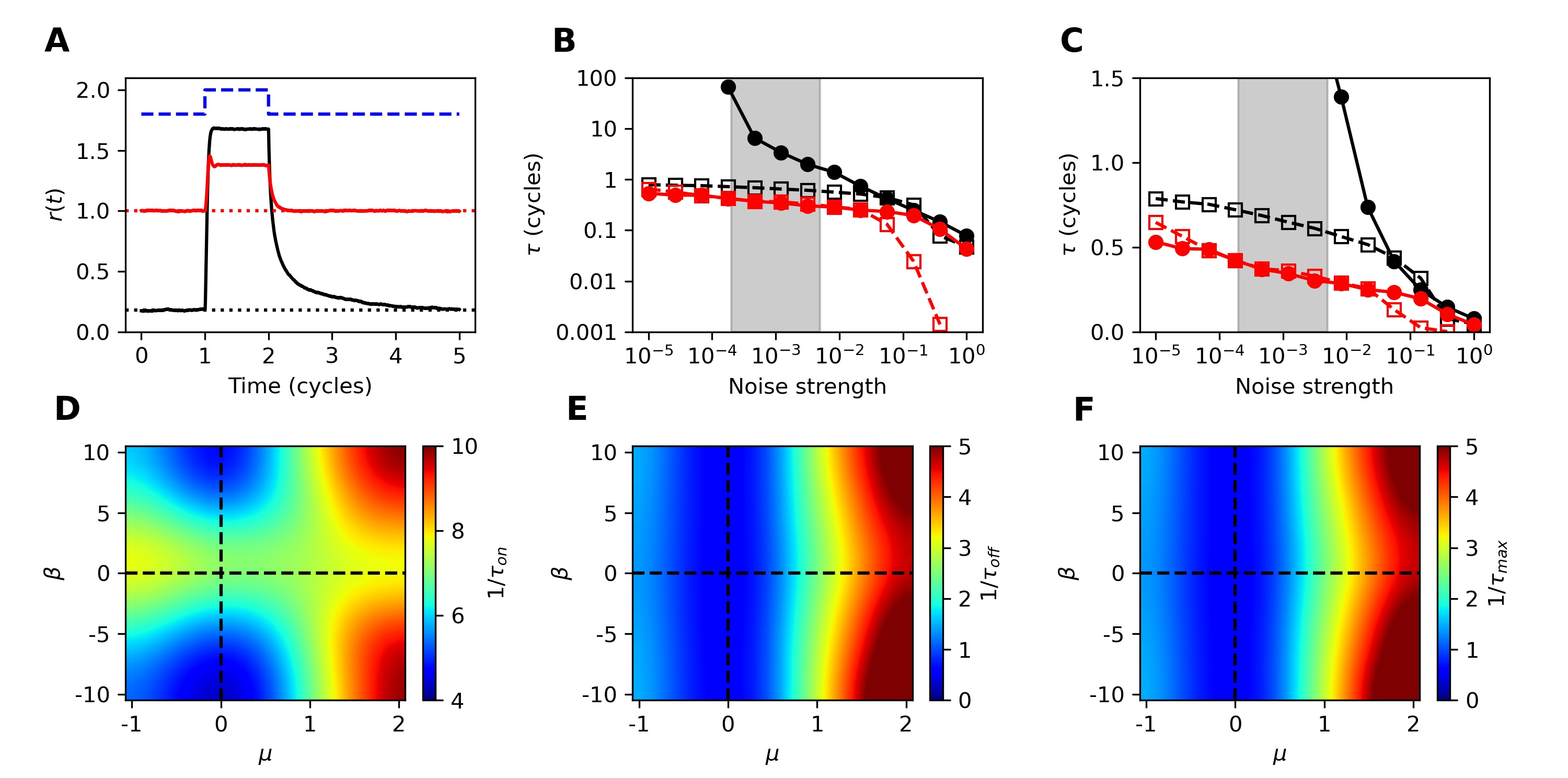}
\caption{Responses of the critical ($\mu =\beta = 0$, black curve) and chaotic ($\mu =1, \beta = 5$, red curve) systems to a step stimulus (illustrated by the blue dashed curve). Response curves represent averages over 64 simulations, each with different initial conditions and realizations of noise. The dotted lines indicate the mean amplitude prior to the step onset. (B) Response times (open squares) and return times (filled circles) of the critical (black) and chaotic (red) systems for several levels of noise. The return time of the critical system rapidly diverges as the noise is reduced. Return times longer than 100 cycles are not plotted. The grey, shaded regions estimate the biological noise level experienced by hair cells. (C) Same data as (B) but zoomed in and on a linear scale. (D-F) Heatmaps of the the speed of response, speed of return, and slower of the two measures. For every combination of parameters, the return time exceeded the response time ($\tau_{off} > \tau_{on}$). For all heatmaps, $D=10^{-3}$.}
\label{fig:Temporal}
\end{figure*}

\section{Mechanical Sensitivity and Information Transfer}

Near the Hopf bifurcation, a noiseless system displays immense sensitivity, with the amplitude gain of the response diverging precisely at the bifurcation \cite{choeModelAmplificationHairbundle1998}. However, this compliance also makes the system susceptible to stochastic fluctuations. In the oscillatory regime, the detector becomes more resistant to noise, but also harder to entrain by external signals. Chaos can assist detectors in synchronizing to external signals; however, this effect can likewise increase susceptibility to external noise and off-resonance stimulus frequencies. We hence determine the sensitivity of the Hopf oscillator throughout the $\mu\beta$-plane and identify the optimal regime for several types of stimulus.

We first consider a weak, single-tone, on-resonance stimulus in the presence of noise. We employ the linear response function, $\chi(\Omega_0)$, to characterize the sensitivity of the system (see Numerical Methods). For systems that exhibit autonomous oscillations, this measure exhibits a spurious non-zero value if the response does not synchronize to the signal. To avoid this issue, we ensure that only the phase-locked component is included in the calculation. We do this by averaging the responses over many systems, each prepared with different initial phases, thereby averaging out any oscillatory component that does not synchronize to the stimulus. It has previously been shown that the isochronous Hopf oscillator detects this signal best when poised as far into the oscillatory regime as possible \cite{maoileidighSinusoidalsignalDetectionActive2018a}. Our results are consistent with this finding, with even further improvement when the system is weakly chaotic (Fig. \ref{fig:Sensitivity}). Further, we show that detectors in this regime outperform those in the critical regime for all levels of noise considered.

Next, we consider stochastic modulations in the amplitude (AM) and frequency (FM) of this signal (see Numerical Methods). Unlike the pure-tone stimulus, these signals carry information in their modulations. We therefore think of the detector not only as a mechanical resonator, but also as an information-theoretic receiver. We employ transfer entropy as the measure of information captured by the receiver \cite{Schreiber2000}. This measure is particularly useful, as it carries no assumptions about which features of the external signal are important. Instead, it directly measures the amount of information transmitted from one process to another, and can even be used to establish causality between two processes. For both types of modulation, we find the oscillatory, chaotic regime to yield optimal detection of information carried by the signal. We also demonstrate the robustness of this measure to stochastic fluctuations and show that this regime is preferred over the critical, non-chaotic regime (Fig. \ref{fig:Sensitivity}).

\begin{figure*}[t]
\centering
\includegraphics[width=\textwidth]{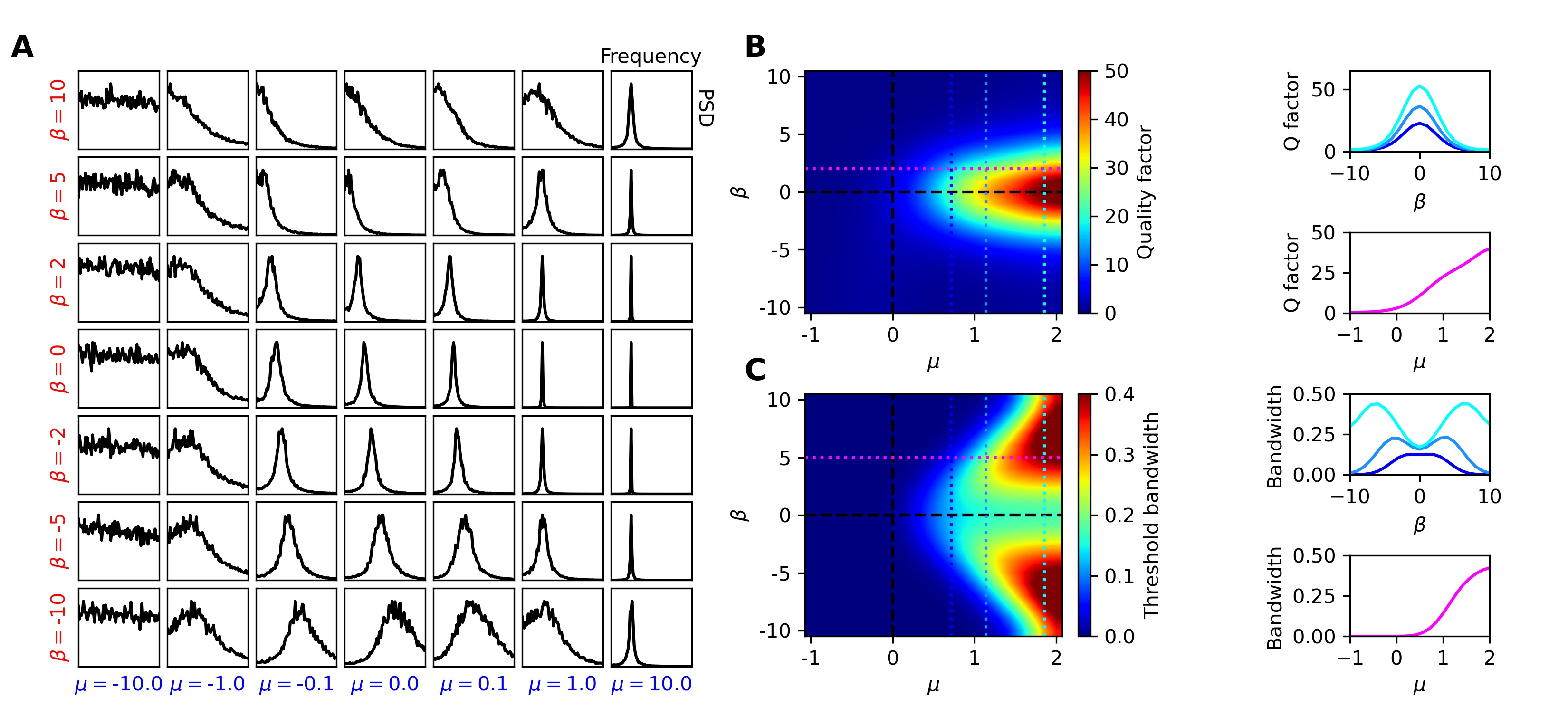}
\caption{(A) Power spectral density of the system through the parameter space, in response to white noise stimulus. All curves are normalized to their peak values and plotted on a linear scale. (B) Heatmap of the quality factor as measured from the response curves. (C) Heatmap of the threshold bandwidth, indicating the frequency range for which the Fourier-component amplitudes of the response exceed $0.1$. For all panels, $D=0.01$. Cross-sections of the heatmaps are shown on the right.}
\label{fig:Tuning}
\end{figure*}

\section{Temporal Acuity}

High temporal acuity is essential for a sensory system to be responsive to brief signals. Further, localization of sound by vertebrates relies on interaural time differences as small as a few microseconds, which correspond to temporal differences of just a fraction of a single stimulus cycle. When such small differences are biologically meaningful, the system's response to a stimulus and its subsequent return to steady state must occur rapidly. To characterize this temporal acuity, we apply a step stimulus and measure the time the system requires to reach its steady state response, as well as the time it takes to return to its unstimulated steady state after cessation of the signal.

We previously found that nonisochronicity, and therefore the degree of chaos, greatly increases the speed of response to a step stimulus. These results are consistent with experimental measurements \cite{faberChaoticDynamicsEnhance2019a}. In the present work, we extend our analysis to variations in the control parameter and determine the temporal acuity of the Hopf oscillator in the $\mu\beta$-plane. We apply a step stimulus to the detector (Fig. \ref{fig:Temporal}A), averaging over many simulations with different initial conditions and realizations of noise. This method of calculating the mean response averages out the stochastic fluctuations, as well as any autonomous oscillations that would otherwise obscure slow modulations to the mean-field response. 

We define the response time ($\tau_{on}$) to be the time it takes the system to settle to and remain within 4 standard deviations of its steady-state mean value following the onset of the step stimulus. Likewise, we define the return time ($\tau_{off}$) to be the time it takes the mean response of the system to become indistinguishable (within a standard deviation) from its value prior to the stimulus. We consider the limiting factor ($\tau_{max}$) to be the maximum of the two time constants. For all parameters tested, the return time was the limiting factor (Fig. \ref{fig:Temporal}D-F). For simplicity, we therefore drop the subscripts and let $\tau = \tau_{max} = \tau_{off}$.

Stochastic fluctuations not only obscure the mean-field response, but can also distort it. As the noise level increases, so does the average amplitude of the system. This increase in baseline amplitude reduces the distance traveled by the returning mean response, effectively increasing the speed ($\frac{1}{\tau}$) according to our definition. We measure the speed over five orders of magnitude in noise strength and find that the chaotic, oscillatory system is faster to return than the critical system (Fig. \ref{fig:Temporal}B). For extremely large levels of noise, the effects of critical slowing down are removed. However, the levels of noise sufficient to equalize the speed of the chaotic and non-chaotic response are so large, leading to a vast reduction in the sensitivity of the system in both regimes (Fig. \ref{fig:Sensitivity}A, C, and E).

\begin{figure*}[t]
\centering
\includegraphics[width=\textwidth]{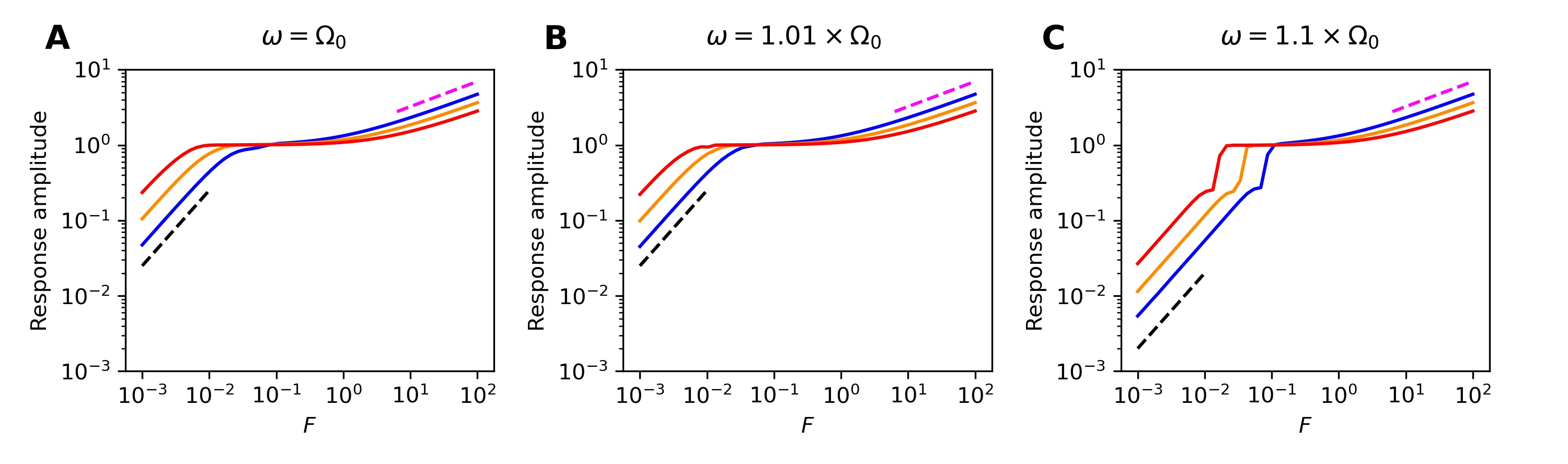}
\caption{Response amplitude from a pure tone stimulus for on-resonance (A) and detuned (B-C) stimulus frequencies as indicated above. Blue, orange, and red curves correspond to $\beta=0$, $2$, and $5$, respectively. Black dashed lines indicate linear growth, while pink dashed lines indicate power-law growth with $|z(\omega)| \propto F^{\frac{1}{3}}$. For all panels, $\mu=1$ and $D=0$.}
\label{fig:Compression}
\end{figure*}

\section{Frequency Selectivity vs Broadband Detection}

Although chaos can make dynamical systems sensitive to small perturbations and external signals, these systems tend to synchronize to a broad range of frequencies. This may be a beneficial effect for broadband detectors, such as vestibular systems, as it would increase the energy transmitted to the system and likely result in lower thresholds of detection. However, for frequency selective detectors, like those of auditory systems, this effect may be harmful. 

We stimulate the Hopf oscillator with additive Gaussian white noise and measure the power spectra of the response throughout the parameter space (Fig. \ref{fig:Tuning}A). As these curves indicate how sharply the system filters the white noise, we use them to characterize the frequency selectivity \cite{Neiman2011}. We observe that the frequency selectivity increases with increasing $\mu$ and decreasing $|\beta|$. Near the Hopf bifurcation, we see that the resonance frequency depends on $\beta$. This is a consequence of the noise altering the steady-state amplitude, and thereby the frequency, in this compliant regime. In the deterministic system, the limit-cycle frequency is independent of $\beta$ by definition ($\Omega_0 = \omega_0 - \frac{\beta\mu}{\alpha} = 1)$. Instead, $\omega_0$ is varied along with the other parameters, so as to keep $\Omega_0$ constant.

To characterize the frequency selectivity of the Hopf oscillator, we employ the quality factor of the response by estimating the full width at half the maximum ($\Delta f$) of the response curves. The unitless quality factor is defined as $Q = \frac{f_0}{\Delta f}$, where $f_0$ is the peak frequency. As expected, the most frequency selective parameter regime occurs at high values of $\mu$ and low values of $|\beta|$ (Fig. \ref{fig:Tuning}B). The quality factor is useful for characterizing the frequency selectivity of a single-frequency or narrowband detector. 

As some auditory and most vestibular systems are responsible for detecting a broader range of frequencies, we utilize the threshold bandwidth ($BW$) for characterizing multi-frequency or broadband detection, a metric suggested in \cite{maoileidighSinusoidalsignalDetectionActive2018a}. We estimate this measure by taking the range of frequencies whose Fourier components have magnitudes exceeding a given threshold. We choose the threshold to be $0.1$, which corresponds to approximately a factor of $10$ above the noise floor of the critical system. The threshold bandwidth increases with increasing $\mu$, as the energy from the spontaneous oscillations amplifies the signal (Fig. \ref{fig:Tuning}C). Further, the threshold bandwidth initially increases with increasing $|\beta|$ due to the broadening of the frequency response curves. However, the $BW$ then diminishes for very large values of $|\beta|$, as the energy becomes so spread out in frequency space that few components exceed the threshold.

\section{Power-law Scaling of Response}

The human auditory system can detect a range of stimulus amplitudes spanning over 6 orders of magnitude in pressure. This process relies on nonlinearities in order to both amplify weak sounds and attenuate loud sounds, thereby protecting the system from damage. These nonlinearities have been measured \textit{in vivo} through otoacoustic emissions and through laser measurements of basilar membrane motion. Further, the attenuation of large amplitudes has been observed \textit{in vitro} on active hair bundles \cite{martinCompressiveNonlinearityHair2001}. At weak forcing, hair cells display a linear response; as the stimulus amplitudes increase, the amplitude response scales as $F^\frac{1}{3}$. The linear and 1/3-power-law responses are reproduced well by the Hopf oscillator, as either the linear or the cubic term dominates in different regimes. Near criticality, the range over which the power law is observed increases. 

We have demonstrated that the nonisochronicity parameter, $\beta$, improves the sensitivity to weak signals. We here show that this parameter also causes stronger attenuation of the response at large stimulus amplitudes. This increased attenuation can be understood by calculating the response in the strong-forcing limit, where the cubic term dominates (see the Appendix). In this limit, the response amplitude is scaled by $\frac{1}{(\alpha^2 + \beta^2)^\frac{1}{6}}$. The combination of these two effects yields an increase in the range of stimulus levels over which amplitude compression is observed and, hence, increases the dynamic range of the system. In Fig. \ref{fig:Compression}, we show that a chaotic oscillator with $\beta=5$ compresses the response amplitude by approximately an order of magnitude more than the isochronous oscillator.

\begin{figure*}[t]
\centering
\includegraphics[width=\textwidth]{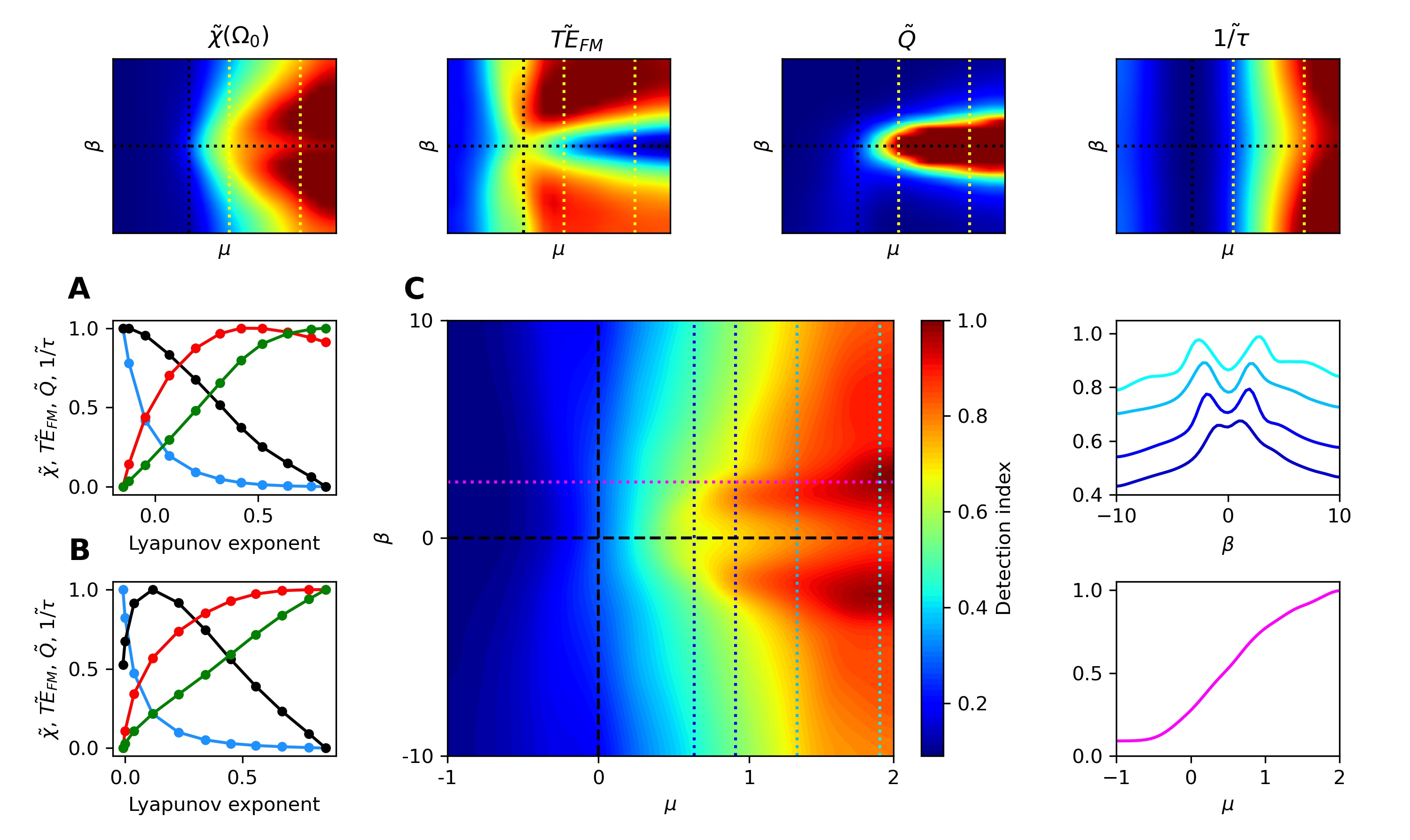}
\caption{(\textit{Optimal Narrowband Detector}) The heatmaps of the four metrics of importance are shown in the top row. The optimal level of chaos for each of the four measures are shown for $\mu\approx 0.5$ (A) and $\mu\approx 1.5$ (B), as indicated by the yellow, dotted lines in the top panels. Black, red, blue, and green correspond to $\tilde{\chi}(\omega_0)$, $\tilde{TE}_{FM}$, $\tilde{Q}$, and $\tilde{\frac{1}{\tau}}$, respectively. (C) Detection index from incorporating the four measures with equal weights, $\mathbf{w} = [\frac{1}{4}, \frac{1}{4}, 0, \frac{1}{4}, 0, \frac{1}{4}]$. Cross-sections of the heatmap at the dotted lines are shown to the right.}
\label{fig:Optimal}
\end{figure*}

\section{Optimal Signal Detector}

The optimal parameter regime depends heavily on the application of the signal detector and the desired specifications. For single-frequency or narrowband detectors, it may be favorable to have the quality factor of the response as high as possible. However, for a broadband detector, a large quality factor would be unfavorable, as it would attenuate frequencies that should be captured. We note that there exists a spectrum of desired capabilities of signal detectors and no single metric can fully characterize the performance. We therefore incorporate all six measures into a single score, which can be weighted in accordance with the application of the system. We first define the vector

\begin{eqnarray}
\mathbf{v} = \begin{bmatrix} \tilde{\chi}(\omega_0), & \tilde{TE}_{FM}, & \tilde{TE}_{AM}, & \tilde{Q}, & \tilde{BW}, & \tilde{\frac{1}{\tau}} \end{bmatrix},
\end{eqnarray}

\noindent where the elements of this vector are the on-resonance linear response, transfer entropy from FM stimulus, transfer entropy from AM stimulus, quality factor, threshold bandwidth, and speed of response, respectively. Each element is linearly scaled such that its range runs from $0$ to $1$ at the minimum and maximum values in the parameter space. To better illustrate the optimal regime, we scale the measures such that they saturate at the 90th percentile of their distributions. We then define the detection index to be the sum of the elements in $\mathbf{v}$, scaled by relative weights of importance,

\begin{eqnarray}
\text{detection index} = \mathbf{w} \cdot \mathbf{v},
\end{eqnarray}

\noindent where $\mathbf{w}$ is a vector whose values represent the relative weights of importance of the measures.

The detection index could optimize at several different places, depending on the desired specifications. From the biological perspective, not only do auditory and vestibular sensors have different demands, but moreover, the same end organs from different species are likely to be optimized for different environments and various conspecific calls. Furthermore, even the same system may self tune into different regimes when placed in different acoustic surroundings. These variations can readily be captured by the appropriate selection of the weighting factor, $\mathbf{w}$. We show a possible selection of weights that illustrates how a narrowband auditory organ may be optimized.

This narrowband detector should prioritize the linear response function at resonance, the quality factor of the response, the speed of response, and the transfer entropy from a narrowband FM signal. Since a large threshold bandwidth would be harmful to this detector's purpose, we set its weight to zero and use the weight vector, $\mathbf{w} = [\frac{1}{4}, \frac{1}{4}, 0, \frac{1}{4}, 0, \frac{1}{4}]$. Using these weights, we find that the optimal regime resides far from the Hopf bifurcation, plateauing at large values of $\mu$, and with a small amount of chaos (Fig. \ref{fig:Optimal}C). For a broadband, vestibular organ, we predict that the optimal regime is in the oscillatory regime, far from the Hopf bifurcation and with a higher degree of chaos (see the Appendix).

\section{Discussion}

We have determined the performance of the Hopf oscillator as a signal detector throughout its parameter space by varying the proximity to criticality and the degree of chaos. To the best of our knowledge, the intersection of these two properties has not previously been studied in the context of optimal signal detection. We first calculated the Lyapunov exponent of the system in the $\mu\beta$-plane, near the Hopf bifurcation, characterizing the level of chaos induced by noise. We showed the breakdown of a previous analytic approximation as the system approaches the Hopf bifurcation. Instead of diverging, the Lyapunov exponent decreases to zero continuously and becomes negative for $\mu < 0$, as is expected for dynamics near a fixed point. This allows us to correlate the degree of chaos with the nonisochronicity parameter, $\beta$.

Providing the oscillator with several types of stimulus, we demonstrated that the sensitivity and temporal acuity of the chaotic system can exceed that of the critical system, regardless of the level of external noise. Further, we showed that the chaotic system compresses the response of the system to large-amplitude signals, simply due to the increase in magnitude of the cubic parameter, $|\alpha + i\beta|$. This increases the dynamic range of stimulus amplitudes the system can detect before damage occurs. All three of these measures (sensitivity, temporal acuity, and amplitude compression) have been shown to be important aspects for signal detection by the auditory and vestibular systems. 

The optimal parameter regime for frequency selectivity, when assessed in isolation,  depends heavily on the application of the detector. The tuning curves broaden as the system approaches criticality from the oscillatory side, and when the level of nonisochronicity is increased. The most sharply-tuned detectors can be found when the system is isochronous and as far into the oscillatory regime as possible. Hence, the only one of our metrics that is degraded by chaos is the quality factor. Note, however, that it is very weakly reduced by small amounts of nonisochronicity. The other five metrics are improved by chaos (Fig. \ref{fig:Optimal}A-B and Fig. \ref{fig:Optimal2}A-B). Temporal acuity monotonically increases with the degree of chaos, while the other metrics (linear response function, transfer entropy, and threshold bandwidth) are maximized at a finite level of chaos.

When considering all measures together, the optimal parameter regime depends on the application of the detector and the desired specifications. We consider two scenarios when proposing optimal regimes. The first is a single-frequency or narrowband detector. When incorporating all of the relevant measures for this detector, we find that the optimal regime is in the oscillatory state, far from the Hopf bifurcation, with weakly chaotic dynamics. We propose that this regime would be well-suited for frequency-selective auditory systems. For the second scenario (see the Appendix), we consider a system designed to detect as broad a range of frequencies as possible, and capture information from transient stimuli, which tend to contain many frequencies. In this case, we show that the optimal regime resides in the oscillatory regime, with a large degree of chaos. We propose that this regime is well-suited for vestibular systems.

Overall, we show that all the relevant detection properties are enhanced by the presence of chaos. Specifically, the issue of critical slowing down is removed, while sensitivity is enhanced. Hence, this tradeoff inherent in critical systems is resolved, and evaluating different regimes of detection - broadband versus frequency selective - changes only the optimal level of chaos. Finally, optimal performance is achieved, not at a specific point, but rather in a broad range of parameter space, which would endow the biological system with flexibility and robustness and obviate the need for extremely precise fine-tuning of parameters.

The current study considers only individual Hopf oscillators, representing single sensory elements. In certain species, the sensory hair bundles are free-standing \cite{manleyOtoacousticEmissionsHair1997}; however, hair bundles of most vestibular and auditory organs studied display some degree of mechanical coupling to each other. The strength and extent of this coupling varies greatly for different specializations of the sensory organ \cite{Mechanotransduction2019}. Considering the diverse configurations of mechanical coupling, we have focused this study on the ability of individual Hopf oscillators to detect signals. We speculate that the performance of the coupled system will correlate with that of its individual elements. However, we point out two issues of individual detectors that have been shown to resolve in the coupled system. The first is the smearing out of the critical point in the presence of noise (Fig. \ref{fig:Hopf}B). It has been shown that, although noise removes criticality in the individual Hopf oscillator, it can be restored in the coupled system \cite{Risler2004}. Second, although chaos deteriorates the frequency selectivity of individual detectors, it can be restored in arrays of coupled detectors, provided that the elements synchronize to each other \cite{faberSynchronizationChaosSystems2021}.

Nonisochronicity is often excluded from numerical models of these sensory systems. It greatly increases the complexity of the system, leading to multi-stability and a chaotic response to various types of signals \cite{faberNoiseinducedChaosSignal2019, faberChaoticDynamicsEnhance2019a}, including white noise. Further, the nonisochronous term in the Hopf oscillator was shown to lead to violation of a generalized version of the fluctuation dissipation theorem, breaking any simple relations between the system's sensitivity to stimulus and susceptibility to stochastic fluctuations \cite{shethViolationGeneralizedFluctuationdissipation2021a}. In the isochronous picture, poising a system near the Hopf bifurcation can be beneficial. The transfer entropy is maximized (Fig. \ref{fig:Sensitivity}), and the system can achieve large, entrained responses to weak signals, provided that the noise is sufficiently weak. However, to make the system robust to noise, it must reside in the oscillatory regime, thereby utilizing the energy of the autonomous motion to amplify the signal. This results in a tradeoff along the control-parameter axis, leading to an optimal value for $\mu$, where the system is close enough to the bifurcation to entrain to a signal, but oscillatory enough to be robust to noise \cite{maoileidighSinusoidalsignalDetectionActive2018a}.

When nonisochronicity and chaos are introduced, the optimal regime moves deeply into the oscillatory regime, while the entrainability can then be controlled by the level of chaos. This nonisochronicity parameter can be tuned to optimize the detection capabilities of the system, with different optimal $\beta$ values existing, depending on the application of the system. We also note that this parameter controls the degree of synchronization in an array of coupled Hopf oscillators \cite{faberChimeraStatesFrequency2021}. Along with the proximity to criticality, we speculate that the degree of chaos is an important characteristic of other biological systems responsible to detecting signals and those that exhibit synchronization of their active components.

\section*{Numerical Methods}

Stochastic differential equations were solved using Heun's method with time steps ranging from $2\pi\times10^{-4}$ to $2\pi\times10^{-3}$.

\subsection*{Response Amplitude \& Linear Response Function}

To determine phase-locked amplitudes we first compute the average response to a sinusoidal stimulus of 64 systems each prepared with different initial phases uniformly spaced across the deterministic limit cycle. This method ensures that any non-synchronized oscillations at the stimulus frequency will average to zero, and only signals that lock to the stimulus will be counted toward the response. We then fit a sinusoid to the mean response with frequency fixed to the stimulus frequency. We define the phase-locked amplitude as the amplitude of this fit. We then compute the linear response function by dividing this response amplitude by the forcing amplitude.

\subsection*{Frequency-Modulated (FM) and Amplitude-Modulated (AM) Stimuli}

The frequency-modulated forcing takes the form

\begin{eqnarray}
F_{FM}(t) = F_0 e^{i\psi(t)},
\end{eqnarray}

\noindent where $\psi(t)$ is the instantaneous phase of the stimulus, and we set $F_0 = 0.1$. The instantaneous stimulus frequency is centered at $\Omega_0$, with additive stochastic fluctuations,

\begin{eqnarray}
\omega(t) = \Omega_0 + \eta_f(t),
\end{eqnarray}

\noindent where $\eta_f(t)$ is low-pass filtered Gaussian white noise (pink noise) with a brick-wall cutoff frequency of $\Omega_0$. We let the standard deviation of this variable equal $0.3\times\Omega_0$. We can then calculate the instantaneous phase of the stimulus,

\begin{eqnarray}
\psi(t) = \int_0^t \omega(t') dt' = \psi(0) + \Omega_0 t + \int_0^t \eta_f(t') dt'.
\end{eqnarray}

\noindent Information production is determined solely by the frequency modulator, $\eta_f(t)$. This method of generating the signal does not influence the amplitude and allows us to examine the effects of information transmission through frequency modulation alone.

Similarly, amplitude-modulated signals take the form

\begin{eqnarray}
F_{AM}(t) = \eta_a(t) e^{i\Omega_0 t},
\end{eqnarray}

\noindent where $\eta_a(t)$ is low-pass filtered Gaussian white noise (pink noise) with a brick-wall cutoff frequency of $\Omega_0$. We set the mean and standard deviation of this stochastic amplitude modulator to be $0$ and $0.5$, respectively.

\subsection*{Transfer Entropy}

The transfer entropy \cite{Schreiber2000} from process $J$ to process $I$ is defined as

\begin{eqnarray}
T_{J \rightarrow I} = \sum p(i_{n+1}, i_{n}^{(k)}, j_{n}^{(l)})
\log \frac {p(i_{n+1} \ | \ i_{n}^{(k)}, j_{n}^{(l)})} {p(i_{n+1} \ | \ i_{n}^{(k)})},
\label{eq:TE}
\end{eqnarray}

\noindent where $i_{n}^{(k)} = (i_n,...,i_{n-k+1})$ are the $k$ most recent states of process $I$. Therefore, $p(i_{n+1} \ | \ i_{n}^{(k)}, j_{n}^{(l)})$ is the conditional probability of finding process $I$ in state $i_{n+1}$ at time $n+1$, given that the previous $k$ states of process $I$ were $i_{n}^{(k)}$ and that the previous $l$ states of process $J$ were $j_{n}^{(l)}$. The summation runs over all points in the time series and over all accessible states of both processes. The transfer entropy measures how much one's ability to predict the future of process $I$ is improved upon learning the history of process $J$. The measure is asymmetric upon switching $I$ and $J$, as information transfer between two processes is not necessarily symmetric. For stimulus-response data, this asymmetry allows for control tests by measuring the transfer entropy from the response to the stimulus, which should be zero. The choice of $k$ and $l$ has little effect on the results, so we select $k = l = 5$, and sample the 5 points such that they span one mean period of the system. We discretize the signal into 4 amplitude bins, however, similar results were obtained when using 2 bins.

\subsection*{Return Time}

We determine the return time, $\tau$, by calculating the average response from 64 simulations to a large-amplitude step stimulus,

\begin{eqnarray}
F(t) = F_{0}[\Theta(t - t_{on}) - \Theta(t - t_{off}) ],
\end{eqnarray}

\noindent where $\Theta(t)$ is the Heaviside step function, and we set $F_{0} = 5$, $t_{on} = 1$, and $t_{off} = 2$. We define the return time as the time it takes the mean response of the system to return to a value within a standard deviation of the mean steady-state amplitude, as measured by the data prior to the step onset.

\subsection*{Quality Factor \& Threshold Bandwidth}

To estimate the quality factor of the system response, we stimulate with additive white Gaussian noise ($D = 0.01$), and calculate the average power spectrum over 60 simulations, each with different initial conditions and realizations of noise. This produces a smooth curve, which can then be used to estimate the full width at half maximum and the quality factor of the response. We then use this curve to calculate threshold bandwidth by determining the range of frequencies for which the Fourier amplitudes exceed a threshold of $0.1$. This threshold was chosen as it is approximately an order of magnitude above the noise floor.

\section*{Acknowledgments}

The authors gratefully acknowledge the support of NSF Biomechanics and Mechanobiology under Grant No.
1916136.

\bibliography{Bibliography}

\section*{Appendix}

\begin{figure*}[t]
\centering
\includegraphics[width=\textwidth]{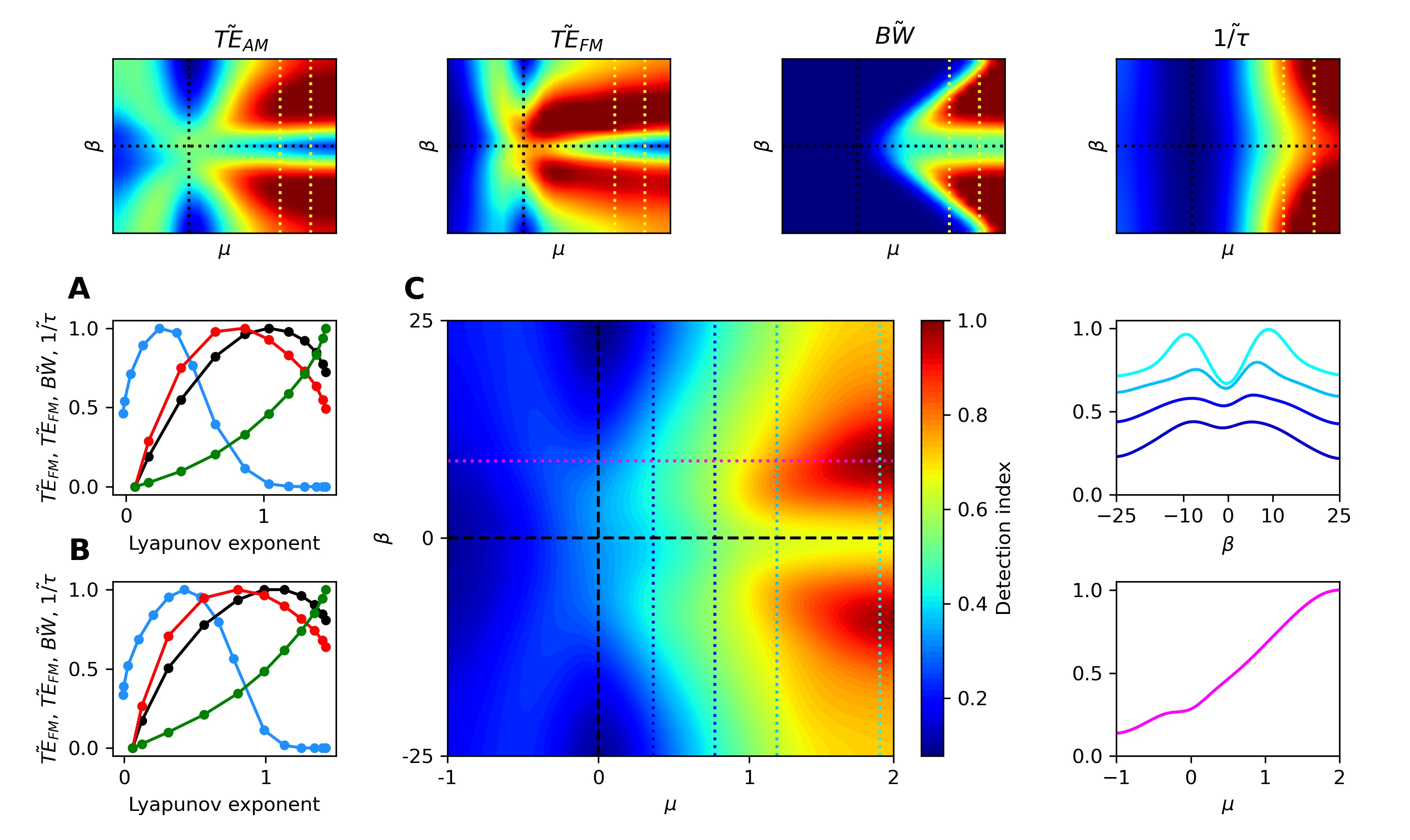}
\caption{(\textit{Optimal Broadband Detector}) The heatmaps of the four metrics of importance are shown in the top row. The optimal level of chaos for each of the four measures are shown for $\mu\approx 1.3$ (A) and $\mu\approx 1.7$ (B), as indicated by the yellow, dotted lines in the top panels. Black, red, blue, and green correspond to $\tilde{TE}_{AM}$, $\tilde{TE}_{FM}$, $\tilde{BW}$, and $\tilde{\frac{1}{\tau}}$, respectively. (C) Detection index from incorporating the four measures with equal weights, $\mathbf{w} = [0, \frac{1}{4}, \frac{1}{4}, 0, \frac{1}{4}, \frac{1}{4}]$. Cross-sections of the heatmap at the dotted lines are shown to the right.}
\label{fig:Optimal2}
\end{figure*}

\subsection{Critical Slowing Down}

We calculate below the time scale associated with transient solutions of the normal form equation, following a weak perturbation or stimulus. At steady-state, the stable solution to Eq. \ref{eq:hopf2b} is $r(t) = r_0 = \sqrt{\frac{\mu}{\alpha}}$ for $\mu > 0$. We find solutions in the vicinity of this limit cycle by letting

\begin{eqnarray}
r(t) = r_0 + \Delta r(t)
\end{eqnarray}

\noindent where $\Delta r(t)$ is the perturbation induced by the stimulus. Plugging this in and utilizing $\frac{\Delta r(t)}{r_0} \ll 1$, we find, to first order, 

\begin{eqnarray}
\frac{d\Delta r}{dt} = -2\mu\Delta r
\end{eqnarray}

\noindent which has solutions that decay with characteristic time

\begin{eqnarray}
\tau = \frac{1}{2\mu}
\end{eqnarray}

\noindent This calculation can also be done in the quiescent regime, which yields $\tau = \frac{1}{\mu}$.

\subsection{Phase-Locking Range}

We consider sinusoidal forcing in the absence of noise ($F(t) = Fe^{i\omega t}$, $D=0$) and express the solution in polar coordinates, $z(t) = r(t)e^{i(\omega t + \phi(t))}$, where $\phi$ is the phase difference between stimulus and response. Eq. \ref{eq:hopf2} becomes

\begin{eqnarray}
\begin{split}
\frac{dr}{dt} = \mu r - \alpha r^3 + F\cos\phi   \;\;\;\;  \text{and} \\   \frac{d\phi}{dt} = \omega_0 - \omega - \beta r^2 - \frac{F\sin\phi}{r}.
\end{split}
\end{eqnarray}

\noindent We consider weak forcing, where $\frac{F}{\mu r_0} \ll 1$ and $r_0 = \sqrt{\frac{\mu}{\alpha}}$. We can then assume that the stimulus only weakly perturbs the amplitude of oscillation, $r(t) = r_0 + \Delta r(t)$ where $\frac{\Delta r}{r_0} \ll 1$. Plugging this in and assuming constant $\Delta r$, we find the steady-state amplitude,

\begin{eqnarray}
r_s = r_0 + \frac{F\cos\phi_s}{2\mu}
\end{eqnarray}

\noindent Inserting this into the phase equation leads to

\begin{eqnarray}
\frac{d\phi_s}{dt} = \Omega_0 - \omega - \frac{F}{r_0}[\sin\phi_s + \frac{\beta}{\alpha}\cos\phi_s] = 0
\end{eqnarray}

\noindent where $\Omega_0 = \omega_0 - \beta r_0$ is the limit-cycle frequency in the absence of stimulus. This equation has solutions for $\phi_s$ only when

\begin{eqnarray}
\frac{F}{r_0} \geq \frac{|\omega - \Omega_0|}{\sqrt{1 + \big(\frac{\beta}{\alpha}\big)^2}}
\end{eqnarray}

\noindent We have thus found the condition for synchronization to the stimulus. Notice that any nonzero value for $\beta$ improves this detectors ability to mode-lock to the external signal. Using this equation, we can also calculate the bandwidth over which the detector will synchronize to the signal:

\begin{eqnarray}
\text{Bandwidth} = 2|\omega_{max} - \Omega_0| = \frac{2F}{r_0}\sqrt{1 + \Big(\frac{\beta}{\alpha}\Big)^2},
\end{eqnarray}

\noindent which increases with increasing $|\beta|$.

\subsection{Nonlinear Response Amplitude}

Beginning with the deterministic, sinusoidally-driven Hopf oscillator,

\begin{eqnarray}
\frac{dz}{dt} = (\mu + i\omega_0)z - (\alpha + i\beta)|z|^2z + F e^{i\omega t},
\end{eqnarray}

\noindent we calculate the steady-state response at the stimulus frequency by assuming $z(t) = R e^{i(\omega t + \phi)}$, where $\phi$ is the phase difference between the response and the stimulus. Plugging this into the differential equation yields,

\begin{eqnarray}
F e^{-i\phi} = (\alpha R^3 - \mu R) + iR(\omega - \omega_0 + \beta R^2).
\end{eqnarray}

\noindent We then multiply by the complex conjugate to find,

\begin{eqnarray}
\begin{split}
F^2 = (\alpha^2 + \beta^2)R^6 + 2\big[(\omega-\omega_0)\beta - \mu\alpha\big]R^4 \\ + \big[\mu^2 + (\omega-\omega_0)^2\big]R^2.
\end{split}
\end{eqnarray}

\noindent In the limit of large $F$, we keep only the leading term in $R$ and find

\begin{eqnarray}
R \approx \frac{F^\frac{1}{3}}{(\alpha^2 + \beta^2)^\frac{1}{6}}.
\end{eqnarray}

\subsection{Optimal Regime (Broadband Detector)}

We now consider a broadband detector, where the threshold bandwidth is important instead of quality factor. The detector should be sensitive to signal modulations and be able to detect transient signals composed of many frequencies. In this case, the information captured by signal modulations is valuable, as well as the speed of response. We therefore choose the weight vector, $\mathbf{w} = [0, \frac{1}{4}, \frac{1}{4}, 0, \frac{1}{4}, \frac{1}{4}]$. This detector is optimized in a manner similar to the narrowband detector in that the ideal regime is deeply into the oscillatory regime. However, a higher degree of nonisochronicity, and therefore chaos, is preferable (Fig. \ref{fig:Optimal2}).

\clearpage

\end{document}